\documentclass[12pt]{article}
\newcommand{\be}{\begin{equation}}
\newcommand{\ee}{\end{equation}}
\newcommand{\OO}{{\cal O}}
\newcommand{\sss}[1]{\mbox{\scriptsize #1}}

\newcommand{\lsim}
{\;\raisebox{-.3em}{$\stackrel{\displaystyle <}{\sim}$}\;}

\newcommand{\GeV}{\unskip\,\mathrm{GeV}}
\newcommand{\MeV}{\unskip\,\mathrm{MeV}}

\newcommand{\real}{{\cal\mbox{Re\,}}}

\newcommand{\J}{{\cal J}}
\newcommand{\I}{{\cal I}}
\newcommand{\M}{{\cal M}}
 
\begin{document}

\pagestyle{empty}
\begin{flushright}
{INLO-PUB 4/99}
\end{flushright}
\vspace*{5mm}
\begin{center}
    {\bf SCREENED-COULOMB ANSATZ FOR THE NON-FACTORIZABLE RADIATIVE CORRECTIONS 
        TO THE OFF-SHELL \boldmath$W^{+}W^{-}$ PRODUCTION}\\
\vspace*{1cm} 
{\bf A.P.~Chapovsky}$^{\dagger)\#)}$\\
\vspace{0.3cm}
Instituut--Lorentz, University of Leiden, The Netherlands
\vspace{0.6cm}\\
and 
\vspace{0.5cm}\\
{\bf V.A.~Khoze}$^{*)}$\\ 
\vspace{0.3cm}
INFN-Laboratori Nazionali di Frascati,
P.O. Box 13, I-00044, Frascati (Roma), Italy\\
\vspace{0.5cm}
\vspace*{2cm}  
                                {\bf ABSTRACT} \\ 
\end{center}
\vspace*{5mm}
\noindent
We demonstrate that the results of the complete first order calculation
of the non-factorizable QED corrections to the single-inclusive cross-sections
for $e^{+}e^{-}\to W^{+}W^{-}\to4$ fermions could be very well reproduced by a 
simple physically motivated ansatz. The latter allows to take into account 
effectively  the screening role of the non-Coulomb radiative mechanisms by
introducing a dampening  factor in front  of the width-dependent part of the known 
first-order Coulomb correction, the so-called screened-Coulomb ansatz.
\vspace*{3cm}\\ 
\begin{flushleft}
February 1999
\end{flushleft}
\noindent 
\rule[.1in]{16.5cm}{.002in}

\noindent
$^{\dagger)}$Research supported by the Stichting FOM.\\
$^{\#)}$ e-mail: chapovsk@lorentz.leidenuniv.nl\\
$^{*)}$ e-mail: khoze@cern.ch
\vspace*{0.3cm}

\vfill\eject

\setcounter{page}{1}
\pagestyle{plain}

\newpage

%
\section{Introduction}
\label{sec:intro}

A precision study of the $W$-boson physics is one of the main objectives 
of the LEP2 programme. New unique possibilities will be 
opened  by a future high-energy electron (muon) collider. This 
physics goal  requires very accurate theoretical knowledge
of the Standard Model predictions for the process 
\be
\label{ee->ww->4f}
        e^{+}e^{-} \to W^{+}W^{-} \to 4\ \mbox{fermions}.
\ee
In particular, the role of the QED~radiative corrections as well as that of the
finite-width effects should be understood in detail~\cite{ww-review}.

It is well known that the instability of the $W$~bosons (the $W$-boson width 
$\Gamma_{W}\approx2.1\GeV$) can  strongly modify the ``stable~W''
results. Special attention should be payed  to the radiative interferences
(both virtual and real) which interconnect the production and the decay stages
of the process (\ref{ee->ww->4f}). In particular, there is a class of 
contributions
corresponding to the so-called 
``charged-particle poles''~\cite{ww-review,fkm-nf,my-nf,my-calc}, which may 
induce strong dependence
of differential distributions on the $W$-boson virtualities. The final-state 
interactions may result in non-factorizable QED radiative corrections to the 
Born cross-section of~(\ref{ee->ww->4f}).
Recall, that the level of suppression of the width-induced effects depends
on the ``degree of inclusiveness'' of the distribution. Thus, for the totally
inclusive cross-section the QED non-factorizable corrections cancel up to the 
terms of $\OO(\alpha\Gamma_{W}/M_{W})$~\cite{fkm-nf,my-nf}. In contrast,
differential distributions could be distorted on the level of $\OO(\alpha)$.
Particular attention should be payed to the threshold region, 
\be
        E=\sqrt{s}-2 M_{W}\sim\OO(\Gamma_{W}).
\ee
Here the instability-induced modification of the Coulomb interaction between
the slowly moving $W$~bosons is especially significant (see for 
details~\cite{fkm-nf,fkmc}). In \cite{fkmc} it is shown that the $W$-boson 
width effects drastically change the on-shell value of the Coulomb correction
even at $E\gg\Gamma_{W}$, but after integration over the invariant masses
of the $W$~bosons the ``stable~$W$'' result is recovered far above the 
threshold region.

Recall, that in the threshold region the Coulomb contribution can be uniquely
separated from the other electroweak corrections. On the contrary, in the
relativistic region it is neither uniquely defined nor gauge invariant. At larger 
$W$-boson velocities $\beta$, the width-induced modifications of the 
differential
distributions caused by other radiative mechanisms (for example, 
intermediate-final 
or final-final state interferences~\cite{fkm-nf,my-calc}) may become just as 
important. These mechanisms may contribute to both factorizable and 
non-factorizable
corrections. It is discussed in~\cite{my-calc} that in the relativistic region a 
cancellation between the different sources of instability takes place. As a result, 
the non-factorizable corrections may vanish and the stable~$W$~result may be 
recovered. In the ultra-relativistic limit, $(1-\beta)\ll1$, such a cancellation 
appears quite naturally. It has, in fact, its origin in the conservation of 
``charged'' currents (see e.g.~\cite{dkt,khoze-stirling,khoze-sjos}).

In the intermediate region, $\beta\lsim1$, which is relevant for the current
LEP2 energy range, an analysis of the non-factorizable corrections to the 
differential cross-sections of $W^{+}W^{-}$ production requires detailed studies.

During the last few years  there has been a significant progress
in our understanding of the radiative 
effects in the off-shell gauge-boson pair production \cite{ww-review}. 
In particular, recently a complete calculation of the non-factorizable 
corrections to the process (\ref{ee->ww->4f}) has been independently performed by 
two groups~\cite{bbc,ddr1,ddr2} (see also \cite{my-calc}). The results 
are in a very good agreement with each other.

In~\cite{khoze-sjos} an attempt has been made to estimate the possible screening 
impact of other radiative interference mechanisms on the Coulomb scenario in the 
relativistic regime. Driven by physical intuition, the consequences of the so-called
dampened- (screened-) Coulomb ansatz have been briefly considered. Within this ansatz
an extra screening factor $(1-\beta)^{2}$ is introduced in front of the 
width-dependent $\arctan$ term in the known first order unstable Coulomb formula 
(see e.g.~\cite{fkmc,fkm-coul}). Such a simple prescription was motivated by a model 
analysis of~\cite{my-calc}, where a simple scenario was considered in which one of 
the $W$~bosons was assumed to be stable. Another example of the cancellation of the 
off-shell effects at relativistic energies has been known for quite a while
(see e.g.~\cite{dkt,kos}). When considering the gluon radiation corresponding 
to the top production and decay at very high energies, one observes that
the width-dependent effects vanish when the emission at the production and decay
stages are added coherently.

An obvious attractiveness of the screened-Coulomb ansatz is that it is very simple 
and readily allows for a transparent physical interpretation. It could be useful as 
well from the point of view of practical applications. It requires, however, a 
special detailed study in order to understand whether this scenario can be taken 
as a realistic plausible baseline.

It is the aim of this paper to perform a detailed comparison of the screened-Coulomb
ansatz with the results of the recent calculations~\cite{bbc,ddr1,ddr2}. It appears 
that this simple prescription provides one with a surprisingly reasonable quantitative
understanding of the screening of the non-factorizable terms at higher energies.
The results of this paper can equally well be applied to the $\gamma\gamma$
initiated processes.

The paper is organized as follows. In Section~\ref{sec:ansatz} some basic formulae
are presented. In Section~\ref{sec:plots} we study numerically several 
characteristic observables. We conclude in Section~\ref{sec:concl}.
Appendix contains a quantitative analysis of the screening effects
basing on the explicit Feynman diagram calculations.

\section{Ansatz for the non-factorizable corrections}
\label{sec:ansatz}

In the Born approximation for the process (\ref{ee->ww->4f}) there are three
(signal) diagrams where two resonant  $W$~bosons are produced and the background
diagrams where, at most, one resonant $W$-boson is formed. The background 
diagrams
are typically suppressed by $\OO(\Gamma_{W}/M_{W})$ 
[$\OO(\Gamma_{W}^{2}/M_{W}^{2})$] with respect to the leading double resonant 
contributions.

The currently most favourable approach adopted for the calculation of the radiative
corrections to the processes involving unstable particles is the so-called 
{\it pole-scheme}~\cite{pole-scheme}. In the double-pole approximation one considers 
the complete off-shell process as a superposition of the production of a pair 
of unstable particles and their subsequent decays. The radiative effects are 
then naturally separated into two groups:
factorizable and non-factorizable. The first type includes radiative corrections
which can be unambiguously attributed either to the production
or to the decay stage of the process. They exhibit simple analytical behaviour 
in the limit $\Gamma_{W}\to0$. The second type corresponds to the radiative 
interconnections between various stages of the process. 

It is instructive to trace the physical origin of such separation not too far from 
the threshold. When 
considering soft photons, $k^{0}=\omega\ll M_{W}$, the production and decay of 
the $W$~bosons can be regarded essentially as point-like processes with a characteristic 
time scale $t_{\sss{char}}\sim1/M_{W}$. However, due to the $W$-decays, various
stages are separated in time by an intervals $\tau\sim1/\Gamma_{W}$. When we average 
over the times between $W$-pair production and the $W$-decays a significant 
interconnection occurs only in the $\omega\lsim\Gamma_{W}$ domain. This results in 
the non-factorizable correction. The contribution to these corrections caused by 
the hard photons is power suppressed (see e.g.~\cite{fkm-nf,kos,dkos}).

When examining the process (\ref{ee->ww->4f}) one distinguishes three energy 
domains:
\begin{itemize}
        \item Threshold region (2) where
                 $W$'s are moving with a small velocity
                with  respect to each other, 
                $\beta\sim\sqrt{\Gamma_{W}/M_{W}}\ll 1$.
        \item Non-relativistic region, $\Gamma_{W}\ll E\ll M_{W}$,
                where the velocity of the $W$'s is still a small parameter, 
                $\beta\ll 1$, but the centre-of-mass energy
                is sufficiently far from threshold.
        \item Relativistic region, $E\sim M_{W}$, where
                velocity of the $W$'s is not a small parameter
                any more, $\beta\sim1$. 
\end{itemize}
Recall, that in the threshold and in the non-relativistic region the main 
contribution to the radiative corrections comes from the Coulomb interaction 
(see~\cite{fkmc,fkm-coul}). All other effects are suppressed by $\OO(\beta)$. 
Near threshold the Coulomb contribution dominates the instability effects. In 
the relativistic region the terms suppressed in the 
non-relativistic region are not small and should be taken into account. The 
explicit calculation of the complete non-factorizable correction performed 
in~\cite{bbc} uses the ``far from threshold'' (FFT) approximation, which assumes 
that $\Gamma_{W}\ll E$. The accuracy of this approximation is 
$\OO(\Gamma_{W}/E)$.
This approximation breaks down in the threshold region, but it is valid in the 
non-relativistic region (far from threshold) and in the relativistic region. 
Note that in the non-relativistic region the calculation of the complete 
non-factorizable correction agrees with the calculation of the off-shell Coulomb
effect within the adopted approximations.

We discuss below a simple ansatz based of the Coulomb result (screened-Coulomb) 
which appears to be in a good agreement with the complete calculation of the 
non-factorizable corrections in the relativistic as well as in the non-relativistic 
region. Of course, one cannot expect that a simple unique prescription exists, 
which would allow to reproduce reasonably well the results of the explicit 
complete calculations of the non-factorizable corrections to the arbitrary 
differential distribution.%
\footnote{When considering the angular distributions of the final-state 
          fermions the general arguments based on the conservation of the 
          ``charged currents'' (see e.g.~\cite{dkt}) may be not applicable, 
          and the screened-Coulomb ansatz could be  irrelevant. The largest 
          discrepancies should be observed near the edges of the kinematic 
          phase space where the corrections are the largest, but the event 
          statistics is very limited. A detailed study of the dependence of 
          the non-factorizable corrections on the fermion angles has been 
          presented in~\cite{bbc,ddr1,ddr2}.}
Below we concentrate on the quantities, which are inclusive with respect 
to all the decay and production angles, such as the invariant mass spectrum of a 
$W$~boson.

Since the calculation of the non-factorizable corrections in~\cite{bbc,ddr1,ddr2}
had been performed in the FFT approximation, we shall remain within the same 
scheme for the screened-Coulomb ansatz. This  means that we shall not consider 
the threshold region here. The reader is reminded that the latter region 
has been studied in detail elsewhere (see e.g.~\cite{fkmc,fkm-coul}). 
For the reference purposes we consider also a model case when the cross-section 
is corrected by the Coulomb effect only. Then the differential distribution 
over an observable $X$ can be written in the following form%
\footnote{We are considering here and in what follows only the first-order 
        Coulomb formulae. As shown in~\cite{khoze-sjos,fkms} the higher order 
        Coulomb effects are practically negligible.}
\be
        \frac{d \sigma_{\sss{``Coul''}}}{d X} = \frac{d \sigma_{\sss{Born}}}{d X}
        \biggl(1+\delta_{\sss{``Coul''}}\biggr),
        \ \ \ 
        \delta_{\sss{``Coul''}} = 
        \delta_{\sss{``Coul''}}^{\sss{on-shell}} +
        \delta_{\sss{``Coul''}}^{\sss{nf}}
\ee
\be
\label{coulomb}
        \delta_{\sss{``Coul''}}^{\sss{on-shell}} = 
        \frac{\alpha\pi}{2 \beta},
        \ \ \ 
        \delta_{\sss{``Coul''}}^{\sss{nf}}
        =
        - 
        \frac{\alpha}{\beta}
        \arctan\Biggl(\frac{M_1^2+M_2^2-2 M_W^2}{2M_W \Gamma_W}\Biggr),
\ee
where $M_{1}$ and $M_{2}$ are the invariant masses of the $W$~bosons, $\beta$
is their on-shell velocity
\be
        \beta = \sqrt{1-4 M_{W}^{2}/s},
\ee
and $d\sigma_{\sss{Born}}/dX$ is the on-shell Born cross-section. This is the 
leading contribution to the radiative correction in the non-relativistic region.  
All other contributions, which 
were neglected, are suppressed by, at least, $\OO(\beta,\Gamma_{W}/E)$. Let us 
emphasise that outside of the threshold region the Coulomb approach is just an 
oversimplified extreme  and, naturally, is not supposed to correspond to the true 
physics.
Note, that throughout this paper the so-called fixed-width scheme is used, where
$\Gamma_{W}$ is the on-shell $W$-boson width.

The two terms in (\ref{coulomb}) have different nature. The first one represents 
the factorizable part of the Coulomb interaction. It is completely the same as the 
familiar Coulomb effect for the stable case. It should be noted again that at high
energies, where $\beta$ is not a small parameter, this correction is of the same 
order as the rest of the radiative corrections, and is not enhanced in any way. 
Typically, the leading contribution coming from radiative corrections goes from 
$\sim\alpha\pi/\beta$ at threshold to $\sim\alpha/\pi$ far from
threshold. 

The second term is the 
non-factorizable part of the Coulomb correction. It arises due to the 
instability effects. It averages to zero when integrated over the invariant 
masses. As discussed in~\cite{fkm-nf,my-nf} (see also~\cite{fkmc}), this is a 
general feature of the non-factorizable corrections.

The physical reason for the separation between the factorizable and non-factorizable 
corrections is rooted in the difference in the characteristic energies and momenta
of the photons responsible for the different terms in (\ref{coulomb}).

In order to gain insight let us consider the diagram with the photon exchange
between the two $W$~bosons.
The denominator of the propagator of the $W$~boson with the 4-momentum 
$p_{1}^{\mu}$
is
\be
        k^{2}+2kp_{1}+D_{1},
        \ \ \
        D_{1} = p_{1}^{2}-M_{W}^{2}+i\Gamma_{W}M_{W}.
\ee
Not too far from threshold for the on-shell (factorizable) part of the Coulomb 
effect photons with energies
$\omega\sim \beta^{2}M_{W}$ and momenta $\,|\vec{k}|\,\sim \beta M_{W}$ 
are essential. It is worth-while to recall that $1/(\beta^{2}M_{W})$
is the typical interaction time between the $W$~bosons, see~\cite{fkmc}.
In such a case $k^{2}$ can not be neglected in the $W$-boson propagator,
contrary to the $\Gamma_{W}M_{W}$ term, see~\cite{fkm-nf,fkmc}. Therefore,
the  Coulomb effect here remains unchanged by the instability of the $W$~bosons.
On the other hand, only the photons with the energies $\omega\sim\Gamma_{W}$ and 
momenta $|\vec{k}|\sim\Gamma_{W}/\beta$ give the leading contribution to the 
off-shell part of the Coulomb effect. Note that $\beta/\Gamma_{W}$ is the 
typical spatial separation between the diverging $W$~bosons~\cite{fkm-coul}.
Far from threshold, at $M_{W}\gg E\gg \Gamma_{W}$, the two regions in the 
photon energy-momentum space are well separated. Because of this fact the effects 
are additive. Near threshold, where $E\sim\Gamma_{W}$, the two regions start to 
overlap, which is precisely the reason why our approach to the calculation of the 
double pole residues becomes invalid.

As has been already mentioned, in the relativistic domain Coulomb correction does 
not account correctly for all the effects. Instead the complete non-factorizable 
corrections are required
\be
        \frac{d \sigma_{\sss{nf}}}{d X} = \frac{d \sigma_{\sss{Born}}}{d X}
        \biggl(1+\delta_{\sss{nf}}\biggr).
\ee
The explicit expressions for $\delta_{\sss{nf}}$~\cite{bbc,ddr1,ddr2} are 
rather lengthy, and for the purposes of this paper there is no need
to present them here. 

Motivated by~\cite{my-calc,khoze-sjos} we would like to check whether the 
complete non-factorizable corrections could be approximated reasonably well 
by a simple ansatz based on the screening of the non-factorizable (off-shell) 
part of the Coulomb effect
\be
\label{ans_1}
        \frac{d \sigma_{\sss{Ans}}}{d X} = \frac{d \sigma_{\sss{Born}}}{d X}
        \biggl(1+\delta_{\sss{Ans}}\biggr),
\ee
where
\be
\label{ans}
        \delta_{\sss{Ans}} = \delta_{\sss{``Coul''}}^{\sss{nf}} (1-\beta)^{2}.
\ee
Non-factorizable corrections distort the Breit-Wigner distribution over 
the invariant mass of the $W$-boson. This results, in particular, in the 
shift of the maximum of the invariant mass distribution. The potential 
importance of this effect is quite transparent since such a shift
may affect the measurement of the mass of the $W$-boson.
It is possible to estimate this shift from the relative non-factorizable 
correction to invariant mass distribution. We will consider specifically
the distribution over the average invariant mass $\bar{M}=(M_{1}+M_{2})/2$.
The standard expression for the linearized shift is
\be
        \Delta \bar{M} =  \frac{1}{8} \Gamma_{W}^{2} 
        \frac{d\delta_{\sss{nf}}(\bar{M})}{d\bar{M}}\Bigl.\Bigr|_{\bar{M}=M_{W}}.
\ee
Basing on the ansatz prescription (\ref{ans_1}) and (\ref{ans}) for the non-factorizable correction
one arrives at the very simple formula for the shift (see also Ref.~\cite{my-calc})
\be
\label{shift}
        \Delta \bar{M} = - \frac{\alpha}{4}\frac{(1-\beta)^{2}}{\beta}\Gamma_{W}.
\ee

In the following Section we shall investigate numerically how this ansatz 
approximates the complete non-factorizable correction to the single-inclusive 
distributions at various energies. In all cases a very good agreement is 
established. We show some specific examples, which illustrate this statement.

\section{Numerical results}
\label{sec:plots}

\begin{figure}[h]
  \unitlength 1cm
  \begin{center}
  \begin{picture}(13.4,15.5)
  \put(-2,14){\makebox[0pt][c]{\boldmath $1+\delta_{\sss{nf}}$}}
  \put(6,8.7){\makebox[0pt][c]{\boldmath $\bar{M} [\GeV]$}}
  \put(3,8.2){\makebox[0pt][c]{\boldmath $\sqrt{s}=172\GeV$}}
  \put(-2,5){\includegraphics{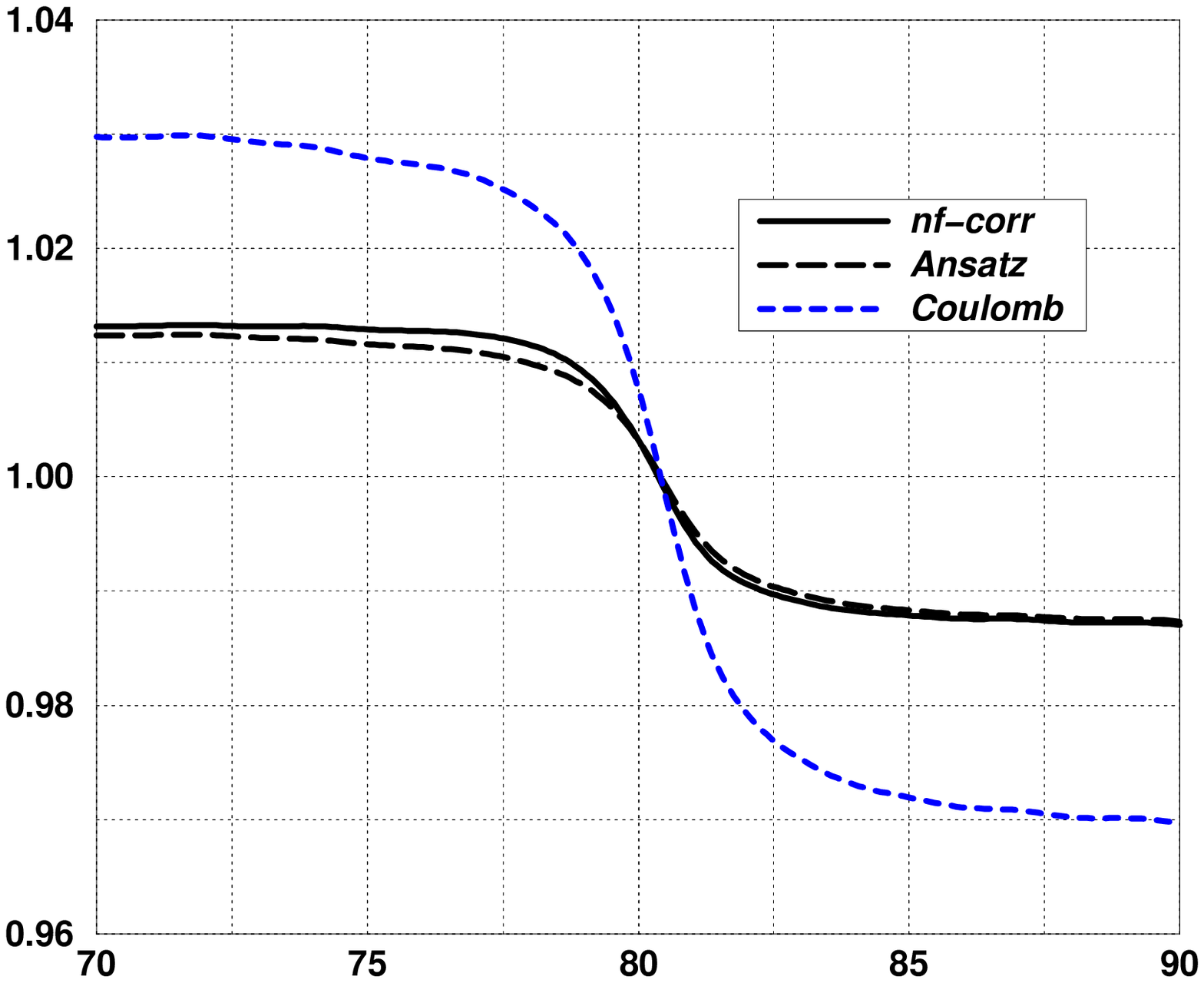}}
  \put(14,8.7){\makebox[0pt][c]{\boldmath $\bar{M} [\GeV]$}}
  \put(11,8.2){\makebox[0pt][c]{\boldmath $\sqrt{s}=183\GeV$}}
  \put(6,5){\includegraphics{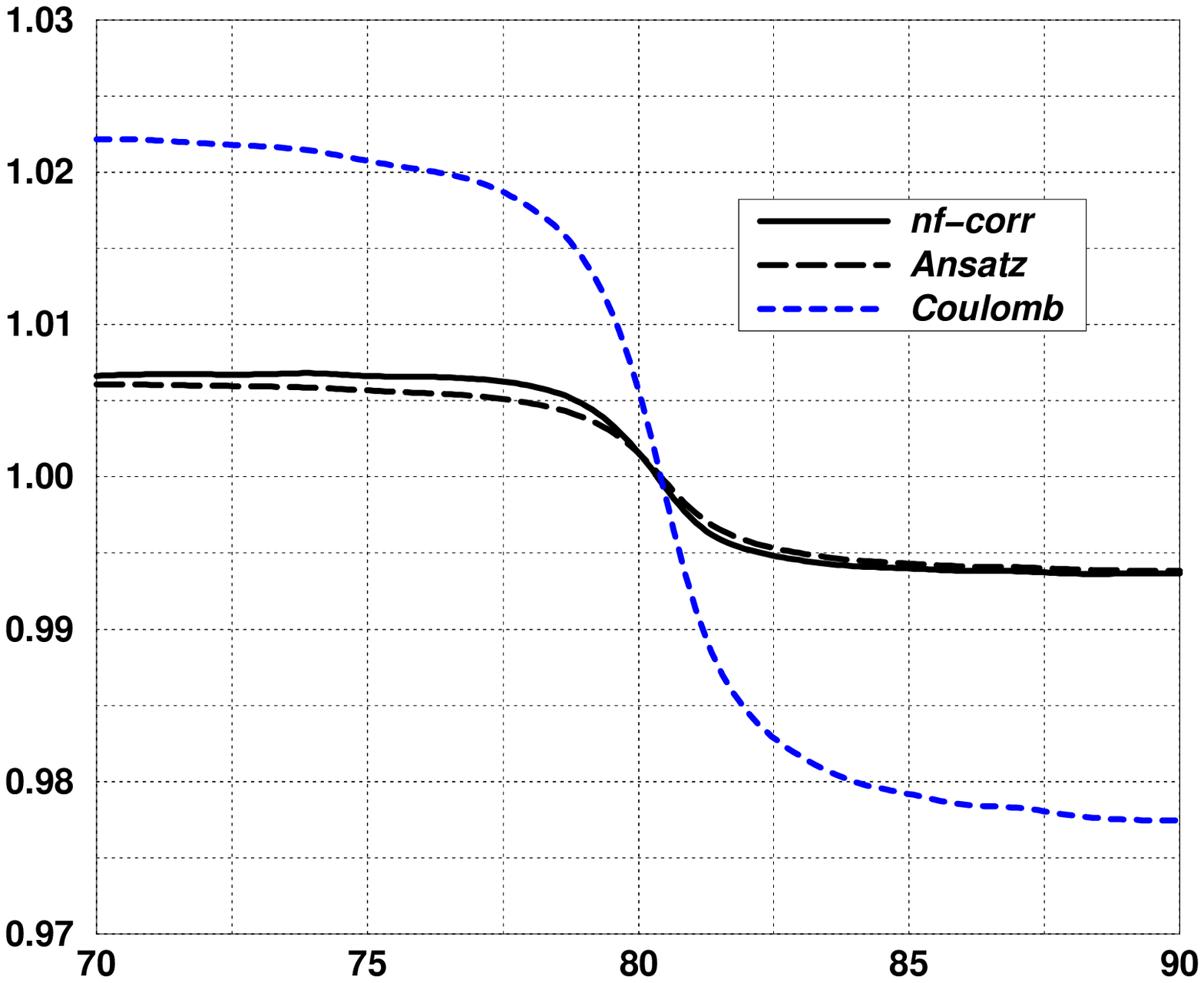}} 
  \put(1.5,6){\makebox[0pt][c]{\boldmath $1+\delta_{\sss{nf}}$}}
  \put(9.5,+0.7){\makebox[0pt][c]{\boldmath $\bar{M} [\GeV]$}}  
  \put(6.5,+0.2){\makebox[0pt][c]{\boldmath $\sqrt{s}=195\GeV$}}
  \put(1.5,-3){\includegraphics{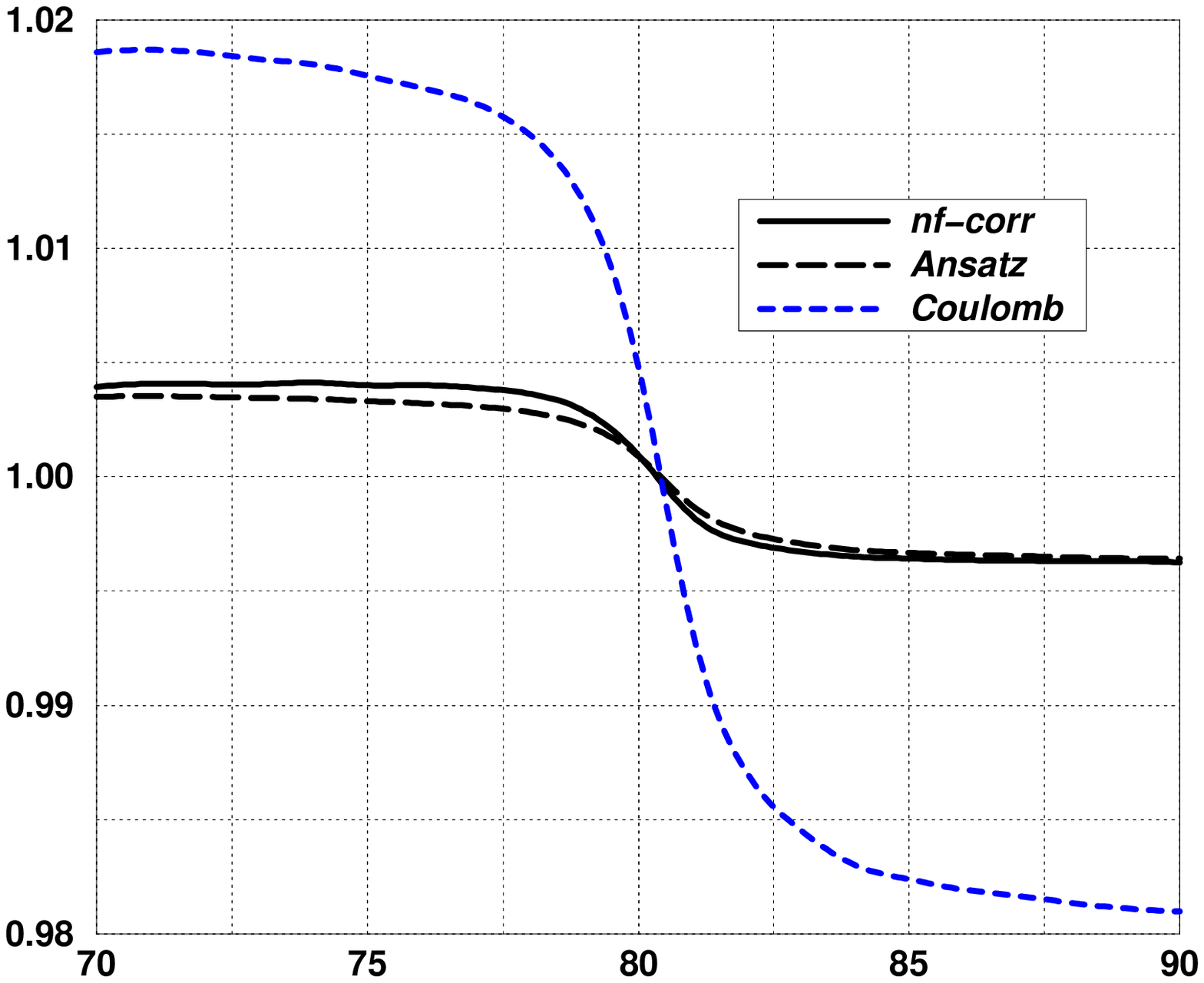}}
  \end{picture}
  \end{center}
  \caption[]{The complete non-factorizable correction 
                to the distribution over 
                the average invariant mass $\bar{M}=(M_{1}+M_{2})/2$ at
                $\sqrt{s}=172$, $183$ and $195\GeV$, as compared to the
                expectations from the screened ansatz and from the 
		unscreened-Coulomb scenarios.}
\label{fig:avg}
\end{figure}%

\begin{figure}[h]
  \unitlength 1cm
  \begin{center}
  \begin{picture}(13.4,6)
  \put(0,4.8){\makebox[0pt][c]{\boldmath $\Delta\bar{M} [\MeV]$}}
  \put(10.5,-0.3){\makebox[0pt][c]{\boldmath $\sqrt{s} [\GeV]$}}
  \put(0,-4.5){\includegraphics{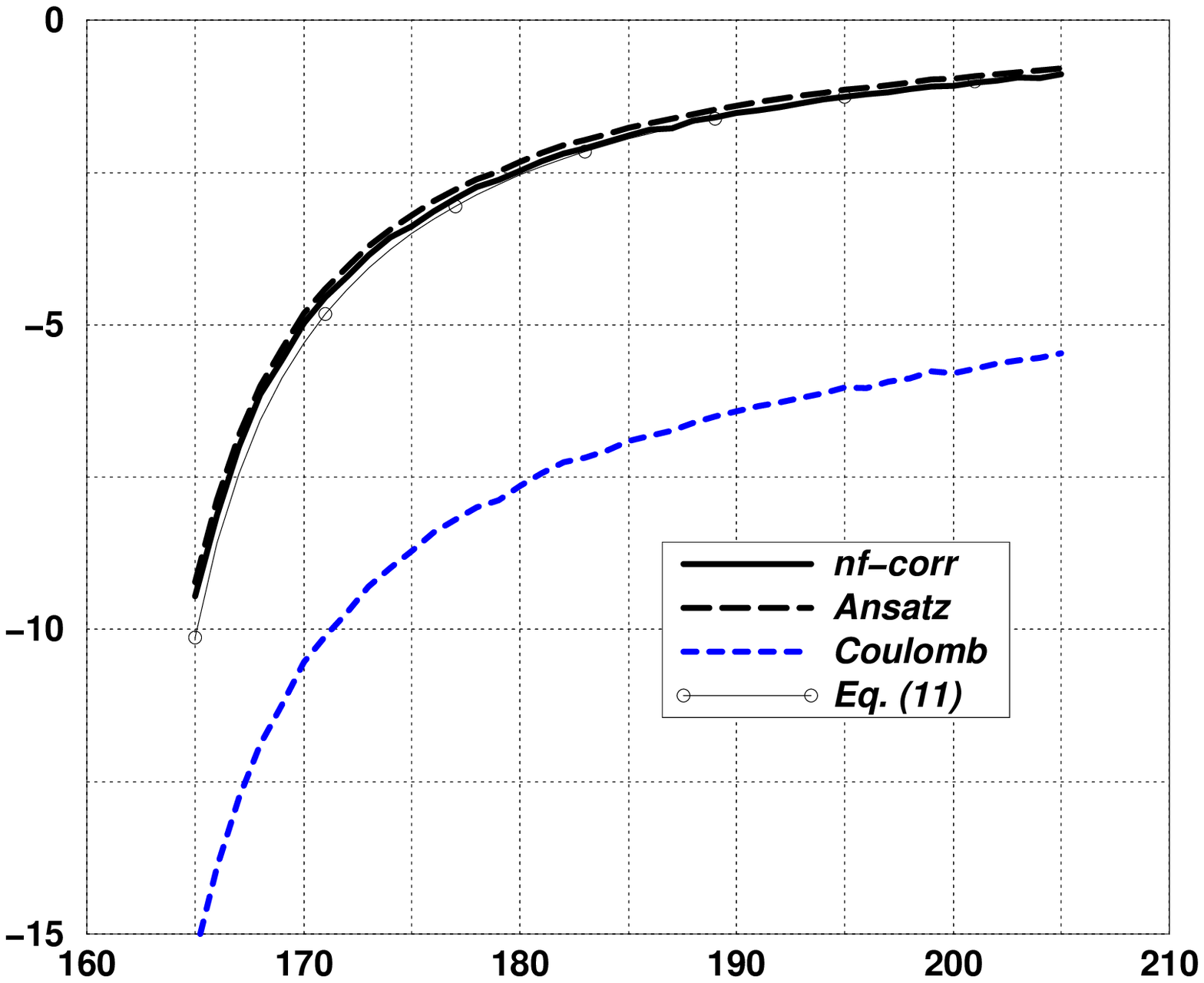}}
  \end{picture}
  \end{center}
  \caption[]{The additional shift of the maximum in the W~average invariant
                mass distribution due to the non-factorizable correction
                as a function 
                of the collider energy, as compared to the expectations from 
		the screened ansatz and from the unscreened-Coulomb 
                scenarios.}
\label{fig:shift}
\end{figure}%

\begin{figure}[h]
  \unitlength 1cm
  \begin{center}
  \begin{picture}(13.4,19)
  \put(-2,14){\makebox[0pt][c]{\boldmath $1+\delta_{\sss{nf}}$}}
  \put(6,8.7){\makebox[0pt][c]{\boldmath $p [\GeV]$}}
  \put(3,8.2){\makebox[0pt][c]{\boldmath $\sqrt{s}=172\GeV$}}
  \put(-2,5){\includegraphics{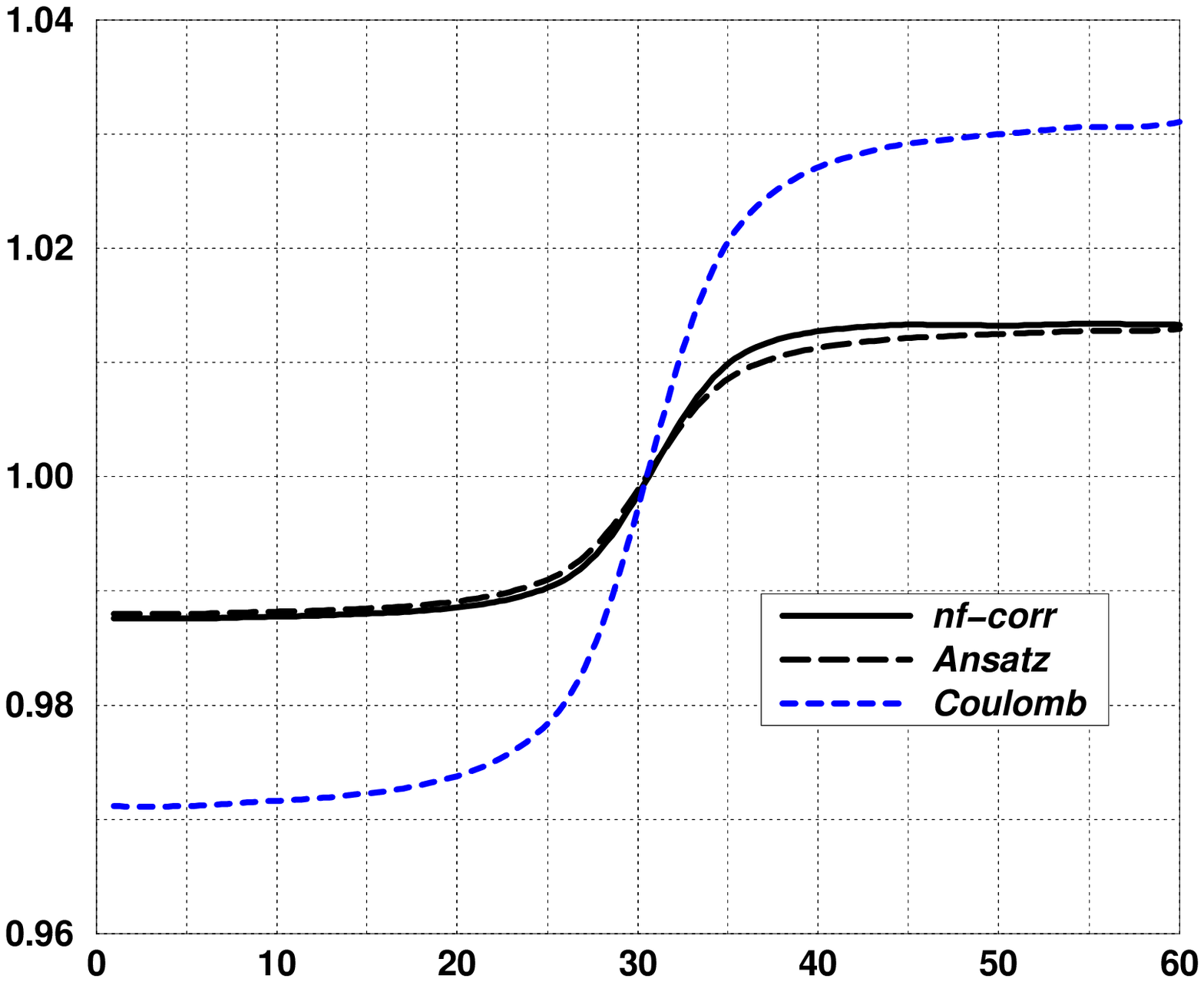}}
  \put(14,8.7){\makebox[0pt][c]{\boldmath $p [\GeV]$}}
  \put(11,8.2){\makebox[0pt][c]{\boldmath $\sqrt{s}=183\GeV$}}
  \put(6,5){\includegraphics{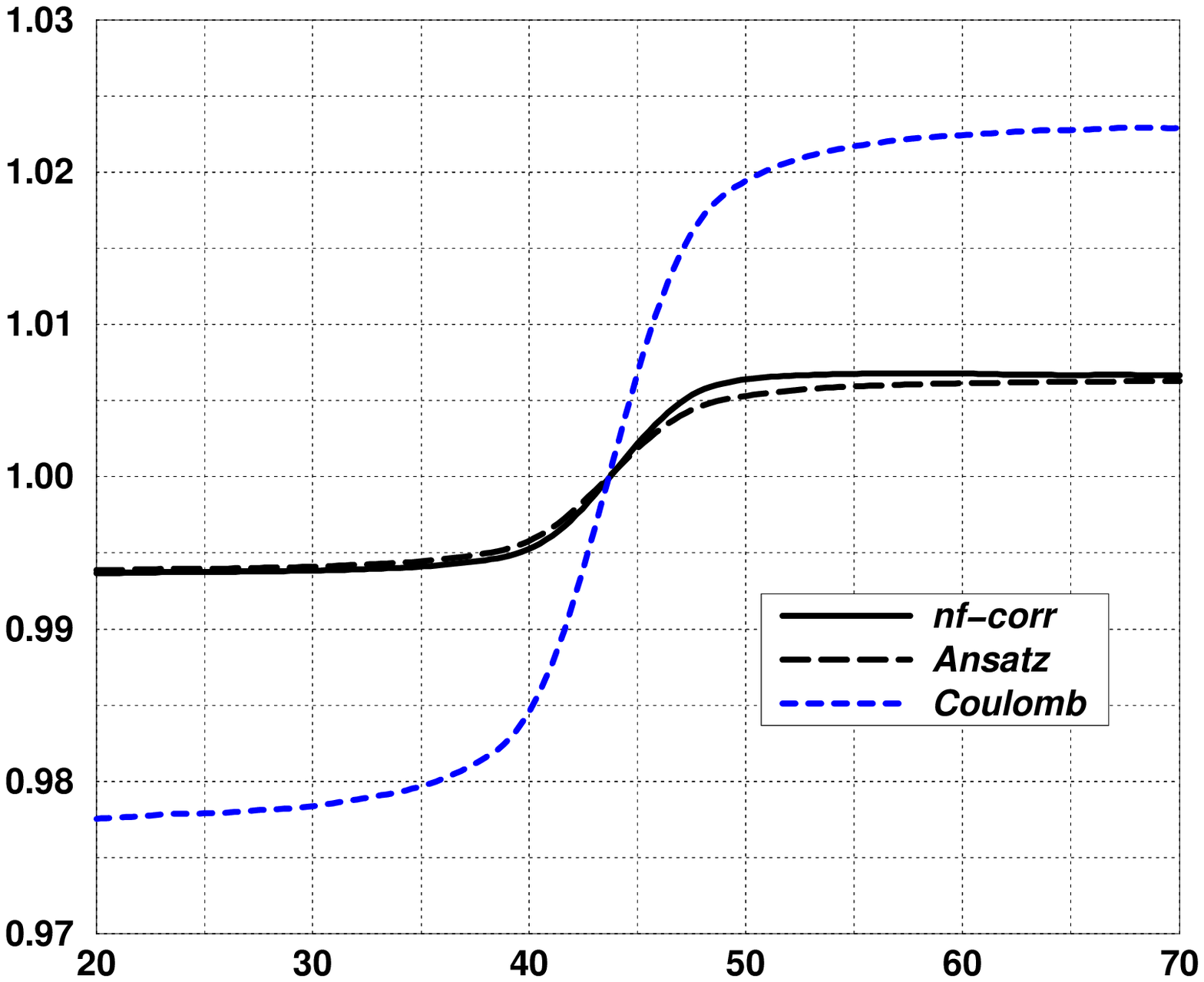}} 
  \put(1.5,6){\makebox[0pt][c]{\boldmath $1+\delta_{\sss{nf}}$}}
  \put(9.5,+0.7){\makebox[0pt][c]{\boldmath $p [\GeV]$}}  
  \put(6.5,+0.2){\makebox[0pt][c]{\boldmath $\sqrt{s}=195\GeV$}}
  \put(1.5,-3){\includegraphics{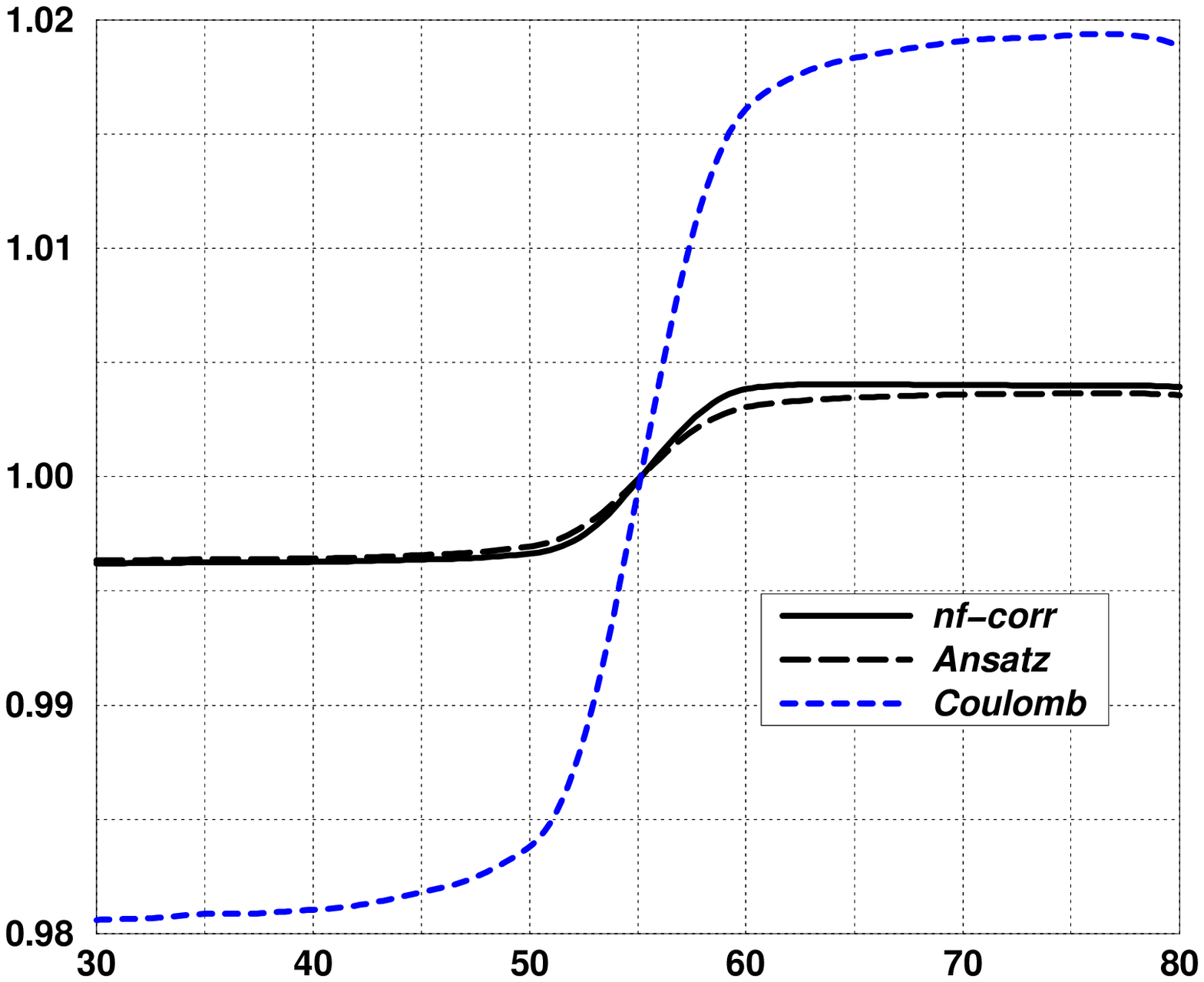}}
  \end{picture}
  \end{center}
  \caption[]{The complete non-factorizable correction
                to the distribution over 
                the $W$-momentum at
                $\sqrt{s}=172$, $183$ and $195\GeV$, as compared to the
                expectations from the screened ansatz and from the 
		unscreened-Coulomb scenarios.}
\label{fig:mom}
\end{figure}%
\begin{figure}[h]
  \unitlength 1cm
  \begin{center}
  \begin{picture}(13.4,6)
  \put(0,4.8){\makebox[0pt][c]{
		\boldmath $(\delta_{\sss{nf}})\biggl[\frac{\sqrt{s}}{2M_{Z}}\biggr]^{4}$}}
  \put(0.3,3.8){\makebox[0pt][c]{\bf [\%]}}
  \put(10.5,-0.3){\makebox[0pt][c]{\boldmath $M [\GeV]$}}
  \put(0,-4.5){\includegraphics{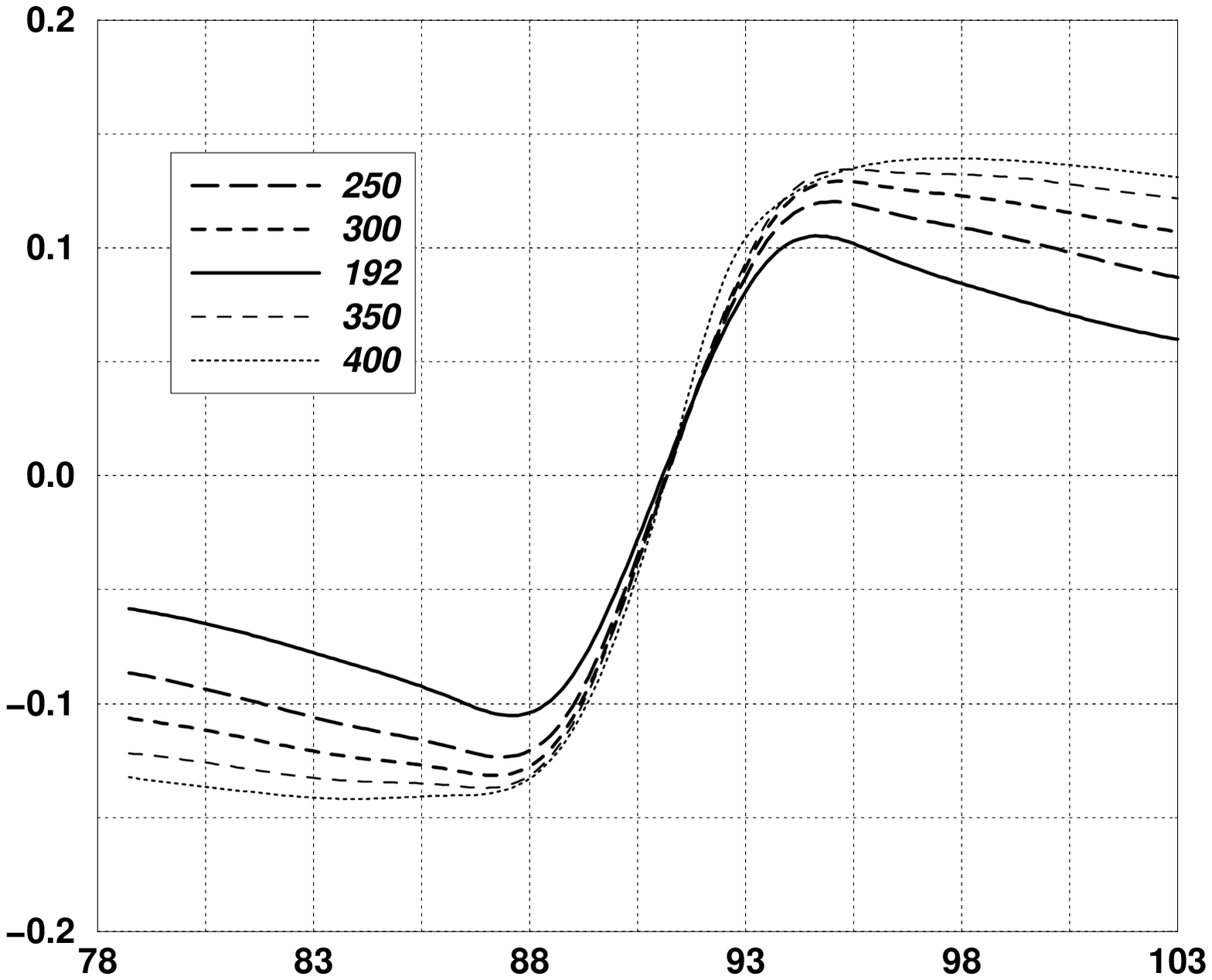}}
  \end{picture}
  \end{center}
  \caption[]{The non-factorizable correction to the distribution
		  over the $Z$-boson invariant mass in 
		  $e^+e^-\to ZZ\to d\bar{d}u\bar{u}$.
		  In order to elucidate the role of the screening factor, the
		  non-factorizable correction, $\delta_{\sss{nf}}$, 
		  was multiplied by
		  the $(\sqrt{s}/2M_Z)^4$ factor.
		  The curves are given for $\sqrt{s}=192$, $250$, $300$, $350$ and $400\GeV$.}
\label{fig:zz}
\end{figure}%
\begin{figure}[h]
  \unitlength 1cm
  \begin{center}
  \begin{picture}(13.4,6)
  \put(0,4.8){\makebox[0pt][c]{\boldmath $\Delta\bar{M} [\MeV]$}}
  \put(10.5,-0.3){\makebox[0pt][c]{\boldmath $\sqrt{s} [\GeV]$}}
  \put(0,-4.5){\includegraphics{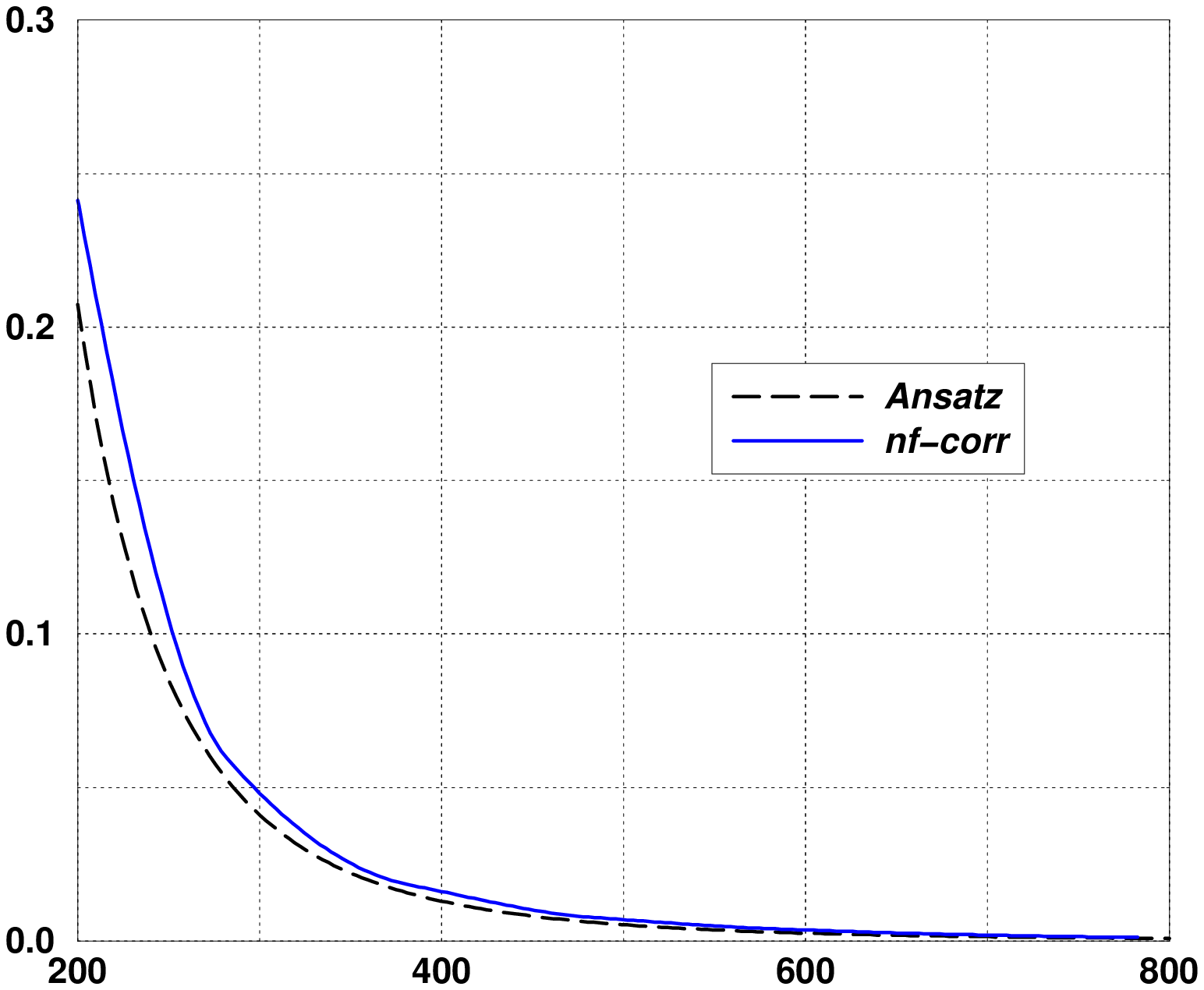}}
  \end{picture}
  \end{center}
  \caption[]{The additional shift of the maximum in the $Z$-boson invariant
                mass distribution due to the non-factorizable correction
                as a function 
                of the collider energy, as compared to the expectations from 
		the screened ansatz scenario:
		$\delta_{\sss{nf}}^{ZZ}\sim (2M_Z/\sqrt{s})^4$.}
\label{fig:zz_shift}
\end{figure}%
In the following calculations we assume
$$
        \alpha=1/137.0359895, \ \ \ 
        \alpha(M_{Z}) = 1/127.9, \ \ \ 
        \sin^{2}\theta_{W}=0.223,
$$
\be
        M_{W}=80.41\GeV, \ \ \    \Gamma_{W}=2.06\GeV,  \ \ \ 
        M_{Z}=91.187\GeV, \ \ \   \Gamma_{Z}=2.49\GeV.
\ee
We use $\alpha(M_{Z})$ to calculate the Born cross-sections,
and $\alpha$ to calculate the radiative corrections.
Results are presented in the  LEP2 energy range $\sqrt{s}=160-200\GeV$ and for 
three discrete energies: $\sqrt{s}=172$, $183$ and $195\GeV$. Several comparisons 
are presented between the results of the complete calculation of the 
non-factorizable corrections, \cite{bbc}, the expectations based on the 
screened-Coulomb 
ansatz, and the model unscreened-Coulomb prescription. The latter scenario 
can help one to assess the impact of the non-Coulomb radiative interferences on 
the width-dependent effects. It should be stressed that throughout the paper
the non-factorizable 
corrections are calculated for the purely leptonic final state 
(for example, $\mu^{+}\nu_{\mu}e^{-}\bar{\nu}_{e}$). Strictly speaking,
outside threshold region
non-factorizable corrections to other final states (i.e. semi-hadronic and purely
hadronic final states) are not identical \cite{ddr2}, but the
differences are, in fact, not so large.

Fig.~\ref{fig:avg} compares the distribution over the average invariant mass 
$\bar{M}=(M_{1}+M_{2})/2$ in the three scenarios  above at 
$\sqrt{s}=172$, $183$ and $195\GeV$. Fig.~\ref{fig:shift} shows the additional
mass shift $\Delta\bar{M}$ due to the non-factorizable effects as a function of the 
collider energy. The expectation corresponding to Eq.~(\ref{shift}) is also shown.
One observes a remarkable agreement between the result of the complete 
calculations and a simple screening recipe (\ref{shift}) for the mass shift.

For practical purposes it is useful to analyse the impact of the instability 
effects on $W$-momentum distribution, see e.g.~\cite{khoze-sjos}. Fig.~\ref{fig:mom}
compares the results for the differential momentum distribution $d\sigma/dp$
in the three scenarios  at  $\sqrt{s}=172$, $183$ and $195\GeV$.

Figs.~\ref{fig:avg} and \ref{fig:mom} clearly show the dampening role
of the screening factor $(1-\beta)^{2}$. In particular, a sharp increase
in $\delta_{\sss{nf}}$ around $p=p_{0}=\sqrt{E M_{W}}$ becomes much
less pronounced as compared to the unscreened case, see also \cite{khoze-sjos}.
The plots demonstrate that the screened-Coulomb ansatz is quite reliable
even for momenta that significantly deviate from $p_{0}$.

Finally, recall that the high energy behaviour of the
non-factorizable corrections to the $ZZ$ production is of
 a special interest, in particular,  because of a certain
resemblance between these QED interference effects and
colour interconnection phenomena in the gauge boson pair
production, see, e.g. \cite{sjos-khoze}. An explicit numerical calculation
 confirms the presence of the same 
screening $(1-\beta)^2$ factor in this case, see Figs.~\ref{fig:zz}
and $\ref{fig:zz_shift}$.
Analogously to the WW case, it also has its origin in the conservation
of currents.

\section{Conclusions}
\label{sec:concl}

The success of the precision studies of the $W$~boson physics relies on an
accurate theoretical knowledge of the details of the production and decay
mechanisms. The instability of the $W$~bosons can, in principle, strongly modify the 
standard ``stable~$W$'' results. An important role can be played by the radiative 
interference effects, which prevent the final state in 
$e^{+}e^{-}\to W^{+}W^{-}\to4$~fermions from being treated as two separate $W$~decays.
Thus, purely QED interaction between two unstable $W$~bosons induces non-factorizable
corrections to various final state distributions. The complete analytical calculations 
of these corrections have been performed only recently \cite{bbc,ddr1,ddr2}. In this
paper we demonstrate explicitly that a simple physically motivated ansatz allows one
to approximate the non-factorizable corrections to the single-inclusive final state 
distributions with a surprisingly good accuracy. This approach makes the physical
insight into the effects of instability of the $W$-pair production quite transparent.

One has to bear in mind that, typically, the order of magnitude of the non-factorizable 
corrections does not exceed one percent, and their practical relevance strongly depends
on the requirements of the experiment. In particular, they could match the expected accuracy
of measurements at a future lepton collider.

Finally, let us note that similar screening scenario could, in principle,
provide a useful framework
for studies of the QCD final-state interactions in $e^{+}e^{-}\to t\bar{t}$
and of the colour interconnection effects in the $W^{+}W^{-}$-production,
see e.g. \cite{sjos-khoze}.
\\
\\
\\
\\
\\
\\
{\bf Acknowledgements}\\
We thank W.~Beenakker, F.~Berends, T.~Sj\"ostrand and W.J.~Stirling for useful discussions.
This work was supported in part by the EU Fourth Framework Programme
'Training and Mobility of Researchers', Network
'Quantum Chromodynamics and the Deep Structure of Elementary Particles',
contract FMRX-CT98-0194(DG 12-MIHT).

\appendix
\renewcommand{\theequation}{\thesection .\arabic{equation}}
\renewcommand{\thesection}{$\!\!$}

\section{ $\!\!\!\!$ Appendix\\
          Screening effects: quantitative discussion}
\setcounter{equation}{0}
\renewcommand{\thesection}{A}

The aim of this Appendix is to expose the origin of the screening $(1-\beta)^2$
factor basing on the explicit evaluation of Feynman diagrams.

We first consider a model case \cite{my-calc} (see also \cite{khoze-stirling}) 
in which one of the W bosons is assumed to be stable and 
the other one has the standard decay modes with decay 
width $\Gamma_W$. In this way we can gain
insight into the analytical structure of the
interference effects without encountering the complications which occur in
the  case of two unstable bosons.
Our final result for the non-factorizable
corrections in such a hypothetical case fully agrees with that in Ref.~\cite{my-calc}.
Nevertheless, we find it instructive to present our alternative (more transparent) 
derivation. The results obtained
within  this simpler model will be used in the 
discussion of the realistic process when both W bosons are off-shell. 

In this model example
the virtual non-factorizable correction can be written as the interference between two 
currents:
\be
\label{my/virt}
         d\sigma_{\sss{nf}} =
         d\sigma_{\sss{Born}}
         2\real 4\pi\alpha i
         \int
         \frac{d^{4}k}{(2\pi)^{4}k^{2}}
         \frac{p_2^{\mu}}{(-p_2 k)}
         \Biggl[\frac{p_1^{\mu}}{p_1 k}-\frac{k_1^{\mu}}{k_1 k}\Biggr] 
         \frac{D_{1}}{D_{1}+2p_{1}k},
\ee
where 
\be
         D_{1}=p_{1}^{2}-M_{W}^{2} + i M_{W} \Gamma_{W},
\ee
$p_{1}$ and $p_{2}$ are the 4-momenta of the unstable $W^{-}$-boson and stable 
$W^{+}$-boson respectively, and $k_{1}$ and $k_{1}'$ are the 4-momenta of the 
decay products of the off-shell $W$-boson, $p_{1}=k_1+k_1'$.

Note, that here and in what follows the $k^{2}$ terms in the propagators
of the radiating particles are neglected. Outside the threshold domain an
account of these terms gives a negligibly small (order $\Gamma_{W}/E$ ) effect.

We must also include the corresponding to (\ref{my/virt}) contribution 
coming from the real photon 
radiatiative interference. Note, that throughout this paper 
we do not consider the non-factorizable corrections involving the initial state
radiation because of the cancellation between the virtual
and real pieces, see for details e.g. the first reference in \cite{fkm-nf}.
This also allows one to apply the results to the photon-photon
initiated process.

The integrand in (\ref{my/virt}) has two poles  in the upper-half $k^{0}$-plane
(the  photon pole and the $W^{-}$-pole). 
The remaining poles are located in the 
lower-half-plane. When we perform the d$k^{0}$ integration by
closing the contour in the upper-half-plane, we see immediately that 
the contributions from the real and
virtual pieces cancel each other. This examplifies the
well-known cancellation between the real and virtual emissions.%
\footnote{This cancellation is similar to the case
          of the non-factorizable corrections caused by the initial-final state 
          radiative interference, see \cite{fkm-nf,my-calc}.}
Thus, the only non-zero contribution comes from the $W^{-}$-pole.
We found it especially
convenient to perform the analysis of this contribution in the rest
frame of the $W^{-}$. In this Lorentz frame the particle 4-momenta can be written as:
          $p_{1}^{\mu}=(E_1; \vec{p}_1)$, 
          $|\vec{p}_1|=\tilde{\beta} E_1$,
          $p_{2}^{\mu}=(M_{W}, \vec{0})$,
          $k_{1}^{\mu} = (\epsilon_{1}; \vec{k}_{1})$

Let us now evaluate  the integral
\be
          I = 
          \real i \int \frac{d^{4}k}{k^{2}}
          \mbox{Pole}^{\sss{up}}
          \frac{(p_2 k_1)}{(-p_2 k) (k_2 k)}\frac{D_{1}}{D_{1}+2p_{1}k}.
\ee 
``Pole$^{\sss{up}}$'' denotes that the residue should be taken in the poles located in the 
upper-half-plane, 
thus only the $(-p_2 k)$-pole contributes. In the $W^{-}$ rest  frame  this term  is
just $-p_2k=-\omega M_{W}+io$. Let us take the residue of  this pole and use the cylindrical 
coordinates
\be
          I = 
          -\real 2\pi  \epsilon_{1} \int
          \frac{dk_{||} k_{\perp} dk_{\perp} d \phi}{k_{||}^{2}+k_{\perp}^{2}}
          \frac{1}{(-k_{||}k_{1 ||} - \vec{k}_{\perp}\vec{k}_{1 \perp} + io)}
          \frac{D_{1}}{D_{1}-2|\vec{p}| k_{||}}.
\ee
Now we carry out the  $k_{||}$ integration.The integrand has radiating
particle poles, which are located in the 
upper-half $k_{||}$-plane. The photon poles occur at $k_{||}=\pm i k_{\perp}$.
Now we close the contour in the lower-half-plane in order to avoid all charged particle poles.
Then the interference contribution becomes
\be
          I = 
          - \real 2\pi^{2}  \epsilon_{1} \int
          \frac{dk_{\perp} d \phi}{k_{\perp}}
          \frac{1}{(ik_{1 ||} - |\vec{k}_{1 \perp}| \cos\phi)}
          \frac{D_{1}}{D_{1}+2|\vec{p}| i k_{\perp}}.
\ee
The integration over the azimuthal angle  $\phi$ is quite straightforward.
The integral over $k_{\perp}$ is infrared divergent, but the divergent piece 
is pure imaginary. Finally, we arrive at
\be
         I\sim \frac{\epsilon_{1}}{|\vec{k_{1}}|} \real i \ln\frac{D_{1}}{i}.
\ee

So far we have evaluated only
the second term in (\ref{my/virt}). The first term  can be treated in an analogous
way after the
substitution $k_{1}\to p_{1}$. The complete non-factorizable correction is then given by
\be
         \delta_{\sss{nf}} \sim 
         \biggl(\frac{E}{|\vec{p}_{1}|}-\frac{\epsilon_{1}}{|\vec{k_{1}}|}\biggr)
         \real i \ln\frac{D_{1}}{i}
         =
         \frac{1-\tilde{\beta}}{\tilde{\beta}}  \real i \ln\frac{D_{1}}{i}.
\ee
The fact that the pre-logarithmic factor approaches zero as $\tilde{\beta}\to1$ is
a direct consequence of the charged current conservation.
Recall, that $\tilde{\beta}=\sqrt{1-M_{W}^2/E_1^2}$ is taken in the system where $W^{-}$ 
is at rest. 
Now we retern to the centre-of-mass frame, where the
velocity of the $W$'s is $\beta$ ($\beta=\sqrt{1-4M_{W}^{2}/s}$).
It is connected to $ \tilde{\beta}$ by $\tilde{\beta}=2\beta/(1+\beta^{2})$.
This allows to present the complete non-factorizable correction in the canonical ansatz form
\be
        \delta_{\sss{nf}}
        \sim
        \frac{(1-\beta)^{2}}{2\beta} \arctan\frac{M_{1}^{2}-M_{W}^{2}}{M_{W}\Gamma_{W}}.
\ee
Now we turn to the realistic case of two off-shell $W$-bosons.
We shall concentrate on the high energy behaviour of the complete
non-factorizable correction.

The virtual non-factorizable correction can be presented in a standard form as a sum
of the current interferences
\be
\label{virt}
 \M_{\sss{nf}}^{\sss{virt}}
 =
 i\M_{\sss{Born}}
 \int\frac{d^{4}k}{(2\pi)^{4}k^{2}}
 \biggl[
 (\J_{0}\J_{+})+
 (\J_{0}\J_{-})+
 (\J_{+}\J_{-})
 \biggr].
\ee
The currents are given by
\be
\label{currents}
 \J_{0}^{\mu}
 =
 e\Biggl[
  \frac{p_{1}^{\mu}}{kp_{1}}
 +\frac{p_{2}^{\mu}}{-kp_{2}}
 \Biggr],
\ \ \ 
 \J_{+}^{\mu}
 = -\,e\Biggl[ \frac{p_{1}^{\mu}}{kp_{1}}
                - \frac{k_{1}^{\mu}}{kk_{1}}
        \Biggr]\frac{D_{1}}{D_{1}+2kp_{1}},
  \ \ \ 
 \J_{-}^{\mu}
 = -\,e\Biggl[ \frac{p_{2}^{\mu}}{-kp_{2}}
                - \frac{k_{2}^{\mu}}{-kk_{2}}
        \Biggr]\frac{D_{2}}{D_{2}-2kp_{2}}
\ee
Here $p_{1,2}$ are the 4- momenta of the $W$-bosons, and $k_{1,2}$ are the 4-momenta
of the corresponding charged decay products. There is also
a corresponding contribution  
coming from the real photon radiation interferences.
The first and the second terms in (\ref{virt})
can be treated exactly in the same way as a model case before.
Therefore, we shall concentrate on the third term which has a different
analytical structure.

Recall, that at higher energies the dominant contribution to the
radiative interference effects comes from the photons (real or virtual)
with the energies 
\be
\label{soft_photons}
           \omega \sim \Gamma_{W} \frac{M_{W}}{E_W},
           \ \ \ 
        2 E_W = \sqrt{s},
\ee
see e.g. \cite{kos}.
One can arrive at the same conclusion from an explicit estimate  of
the dominant contribution to the integral  (\ref{currents}).

To be specific we concentrate below on the typical case of the W decay mass
distribution. Therefore, it is assumed that the integration over the decay products
has been already carried out. Let us analyze the consequences of this
integration for  the Born decay cross-section and for the non-factorizable currents 
(\ref{currents}). First, recall that 
the Born decay matrix element squared can be written as
$$
          \M_{\sss{dec}}^{\mu}\M_{\sss{dec}}^{* \ \nu}
          \sim
          \frac{1}{4}
          \mbox{\bf Sp}
          \biggl[
          \gamma^{\mu} (1-\gamma^{5}) \not k_{1} \gamma^{\nu} (\not p_{1}- \not k_{1})
          \biggr]
          =
          \Delta^{\mu\nu}_{V}-i\Delta_{A}^{\mu\nu}.
$$        
\be
\label{app:born1}
          \Delta_{V}^{\mu\nu} = k_1^{\mu}p_{1}^{\nu} + k_1^{\nu}p_{1}^{\mu}
          -2 k_1^\mu k_1^\nu - g^{\mu\nu}\frac{M_{W}^{2}}{2},
           \ \ \ 
           \Delta_{A}^{\mu\nu} = \epsilon^{\mu\nu k_{1}p_{1}},
\ee
where indices $\mu$ and $\nu$ are to be contracted with the corresponding
ones in the production part of the Born cross-section.
We used the notations $\epsilon^{\mu\nu\rho \, p}=\epsilon^{\mu\nu\rho\alpha} p_{\alpha}$.
Let us start from the vector piece $\Delta^{\mu\nu}_{V}$. 

The Born decay cross-section integrated over the phase-space of the decay products is given by
\be
\label{app:born2}
          \I^{\mu\nu}_{\sss{Born}} =
          \int d^{4} k_{1} \ \delta(k_{1}^{2}) \ \delta\bigl(M_{W}^{2}-2(p_1 \cdot k_1)\bigr)
          \ \times \ 
          \Delta_{V}^{\mu\nu}=
          \frac{\pi}{6} M_{W}^{2} 
          \biggl[g^{\mu\nu}-\frac{p_{1}^{\mu}p_{1}^{\nu}}{M_{W}^{2}}\biggr].
\ee
Consider now an integral over the corresponding non-factorizable current
\be
          \I^{\mu\nu\alpha}_{\sss{nf}, \ V} =
          \int d^{4} k_{1} \ \delta(k_{1}^{2}) \ \delta\bigl(M_{W}^{2}-2(p_1 \cdot k_1)\bigr)
          \ \times \ 
          \Delta_{V}^{\mu\nu}
          \ \times \ 
          \Biggl[ \frac{p_{1}^{\alpha}}{kp_{1}}
                - \frac{k_{1}^{\alpha}}{kk_{1}}\Biggr].
\ee
Tensor $\I^{\mu\nu\alpha}_{\sss{nf}, \ V}$ can depend only on the 4-vectors
$p_{1}^{\mu}$ and $k^{\mu}$ and has the following general features:
\begin{eqnarray}
  &\bullet&    \I^{\mu\nu\alpha}_{\sss{nf}, \ V}=\I^{\nu\mu\alpha}_{\sss{nf}, \ V},
\nonumber\\ 
  &\bullet&    \I^{\mu\nu\alpha}_{\sss{nf}, \ V} p_{1,\ \mu}=0, \nonumber \\
  &\bullet&    \I^{\mu\nu\alpha}_{\sss{nf}, \ V} k_{\alpha} = 0, \nonumber \\
  &\bullet&    \I^{\mu\nu\alpha}_{\sss{nf}, \ V} g_{\mu\alpha}=0. 
\end{eqnarray}
It is convenient to carry out the integration in the $W$-boson rest frame.
The integral $\I^{\mu\nu\alpha}_{\sss{nf}, \ V}$ simplifies in the high energy limit,
if one recalls that only the soft photons (\ref{soft_photons}) are responsible 
for the non-factorizable correction. 
Then $p_{1}^{\mu}\sim E_{W}$, and $k^{\mu}\sim\Gamma_{W} M_{W}/E_{W}$.
\be
          \I^{\mu\nu\alpha}_{\sss{nf}, \ V} = 
          A
          \biggl(\frac{p_{1}^{\alpha}}{k p_1}- \frac{k^{\alpha}}{k^{2}}\biggr)
          \Biggl[g^{\mu\nu}
                 +\frac{k^{2}}{(kp_{1})^{2}} p_{1}^{\mu}p_{1}^{\nu}
                 +M_{W}^{2} \frac{k^{\mu}k^{\nu}}{(kp_1)^{2}}
                 -\frac{1}{(kp_1)}
                  \biggl[k^{\mu}p_{1}^{\nu}+p_{1}^{\mu}k^{\nu}\biggr]
          \Biggr],
\ee
\be
          A = \frac{\pi}{4} \frac{M_{W}^{4}k^{2}}{(kp_1)^{2}}
          \biggl[1+\frac{1}{2}\ln\frac{M_{W}^{2}k^{2}}{4(kp_{1})^{2}}\biggr].
\ee
Note that $A\sim E_W^{-2}$ (we keep track of the energy dependence only),
since $(kp_{1})\sim M_{W}\Gamma_{W}$ and $k^{2}\sim \Gamma_{W}^{2} M_{W}^{2}/E_W^{2}$.
Recall also that the $\mu\nu$ tensor in the equation above is of the same order 
(or less) as in the Born approximation, see (\ref{app:born2}).

Now we can readily obtain an upper limit for the third term in the square brackets in
the integrand in (\ref{virt}).
\begin{eqnarray}
      \M_{\sss{nf, \ V}}^{\sss{virt}}
&       \sim
       \int d^{4} k_{1} \ \delta(k_{1}^{2}) \ \delta\bigl(M_{W}^{2}-2(p_1 \cdot k_1)\bigr)
          \ \times \ 
          \Delta_{V}^{\mu\nu}(p_1, k_1) 
          \ \times \ 
& \nonumber \\
&       \ \times \ 
       \int d^{4} k_{2} \ \delta(k_{2}^{2}) \ \delta\bigl(M_{W}^{2}-2(p_2 \cdot k_2)\bigr)
          \ \times \ 
          \Delta_{V}^{\mu'\nu'}(p_2, k_2)   \ \times \ 
&
       \int\frac{d^{4}k}{(2\pi)^{4}k^{2}}
       (\J_{+}\J_{-})
       \lsim \nonumber
\end{eqnarray}
\be 
\label{estimate}
       \lsim
       \biggl[g^{\mu\nu}-\frac{p_{1}^{\mu}p_{1}^{\nu}}{M_{W}^{2}}\biggr]
        \biggl[g^{\mu\nu}-\frac{p_{2}^{\mu'}p_{2}^{\nu'}}{M_{W}^{2}}\biggr]
         \ \times \ 
       \frac{E_W^{-4}}{E_W^{-2}} \cdot E_W^{-2} E_W \cdot E_W^{-2} E_W.
\ee
As a result, the non-factorizable correction is shown to acquire at high energies
an additional (screening) factor 
\be
\label{app:ansatz}
       \delta_{\sss{nf}}\sim \frac{1}{E_W^{4}} \sim (1-\beta)^{2}. 
\ee

To make the consideration complete we turn now to  the axial piece of the Born
decay cross-section, $\Delta^{\mu\nu}_{A}$ term in (\ref{app:born1}).
It is possible to treat this contribution in an analogous way as before.
Here we present the result of the integration over
the decay phase-space
$$
          \I^{\mu\nu\alpha}_{\sss{nf}, \ A} =
          \int d^{4} k_{1} \ \delta(k_{1}^{2}) \ \delta\bigl(M_{W}^{2}-2(p_1 \cdot k_1)\bigr)
          \ \times \ 
          \epsilon^{\mu\nu k_{1}p_{1}}
          \ \times \ 
          \Biggl[ \frac{p_{1}^{\alpha}}{kp_{1}}
                - \frac{k_{1}^{\alpha}}{kk_{1}}\Biggr]
                =
$$
\be
\label{aaa}
            = 
         B 
         \biggl[
         k^{\alpha} \epsilon^{\mu\nu k p_{1}}-
         k^{2}\epsilon^{\mu\nu\alpha p_{1}}\biggr] 
         +
         C 
         \biggl[
         p_{1}^{\alpha} \epsilon^{\mu\nu k p_{1}}-
         (kp_{1})\epsilon^{\mu\nu\alpha p_{1}}\biggr],
\ee
where
\be
         B = -\frac{\pi}{8}\frac{M_{W}^{4}}{(kp_{1})^{3}}
         \biggl[1 + \ln\frac{M_{W}^{2}k^{2}}{4(kp_{1})^{2}}\biggr],
         \ \ \ 
         C = +\frac{\pi}{8} \frac{M_{W}^{2}}{(kp_{1})^{2}}
         \biggl[1 + 
         \frac{3 k^{2} M_{W}^{2}}{2 (kp_{1})^{2}}\ln\frac{M_{W}^{2}k^{2}}{4(kp_{1})^{2}}
         \biggr].
\ee
Note that $B\sim E^{0}$ and $C\sim E^{0}$ at high energy 
(again we keep track of the energy dependence only). However, the Lorentz structure of 
(\ref{aaa}) is different from that in the Born approximation.
Non-factorizable current integrated over the angles of the decay products
$\I^{\mu\nu\alpha}_{\sss{nf}, \ A}$ is supressed by one power of energy
as compared to the Born approximation, $\sim E^{-1}_{W}$. Estimate similar to 
(\ref{estimate}) shows that the relative non-factorizable correction
behaves as $\sim E^{-4}_{W}$, where one half of the supression comes from 
the phase-space of the photon, $\sim E^{-2}_W$, and another one comes from 
the interference between  two non-factorizble currents, $\sim E^{-1}_W$ from each one.
Thus, the result  (\ref{app:ansatz}) remains valid when the axial contribution
is taken into account.

\end{document}